\documentclass[12pt]{article}
\usepackage[a4paper]{geometry}
\usepackage[myheadings]{fullpage}
\usepackage{fancyhdr}
\usepackage{authblk}
\usepackage{graphicx, wrapfig, subcaption, setspace, booktabs}
\usepackage[T1]{fontenc}
\usepackage[font=small, labelfont=bf]{caption}
\usepackage{fourier}
\usepackage[protrusion=true, expansion=true]{microtype}
\usepackage[english]{babel}
\usepackage{url}
\usepackage{sectsty}
\usepackage{amsmath,amsfonts,amssymb,verbatim}
\usepackage{tocloft}
\graphicspath{ {figures/} }
\usepackage{array}

\begin{document}

\title{ \normalsize 
		\LARGE \textbf{Use of Expected Utility (EU) to Evaluate Artificial Intelligence-Enabled Rule-Out Devices for Mammography Screening}
  }

\date{}

\author[1]{Kwok Lung Fan, PhD}
\author[1]{Yee Lam Elim Thompson, PhD}
\author[1]{Weijie Chen, PhD}
\author[2]{Craig K. Abbey, PhD}
\author[1]{Frank W. Samuelson, PhD \thanks{Corresponding author}}
\affil[1]{U.S. Food and Drug Administration}
\affil[2]{Department of Psychological and Brain Sciences, UC Santa Barbara}

\maketitle
\newpage

\sectionfont{\scshape}


\section*{Abstract}

\noindent\textbf{Background}: An artificial intelligence (AI)-enabled rule-out device may autonomously remove patient images unlikely to have cancer from radiologist review. Many published studies evaluate this type of device by retrospectively applying the AI to large datasets and use sensitivity and specificity as the performance metrics. However, these metrics have fundamental shortcomings because they are bound to have opposite changes with the rule-out application of AI.

\noindent\textbf{Method}: We reviewed two performance metrics to compare the screening performance between the radiologist-with-rule-out-device and radiologist-without-device workflows: positive/negative predictive values (PPV/NPV) and expected utility (EU). We applied both methods to a recent study that reported improved performance in the radiologist-with-device workflow using a retrospective U.S. dataset. We then applied the EU method to a European study based on the reported recall and cancer detection rates at different AI thresholds to compare the potential utility among different thresholds.

\noindent\textbf{Results}: For the U.S. study, neither PPV/NPV nor EU can demonstrate significant improvement for any of the algorithm thresholds reported. For the study using European data, we found that EU is lower as AI rules out more patients including false-negative cases and reduces the overall screening performance.

\noindent\textbf{Conclusions}: Due to the nature of the retrospective simulated study design, sensitivity and specificity can be ambiguous in evaluating a rule-out device. We showed that using PPV/NPV or EU can resolve the ambiguity. The EU method can be applied with only recall rates and cancer detection rates, which is convenient as ground truth is often unavailable for non-recalled patients in screening mammography.

\newpage
\section*{Highlights}
\begin{itemize}
    \item Sensitivity and specificity can be ambiguous metrics for evaluating a rule-out device in a retrospective setting. PPV/NPV can resolve the ambiguity but require the ground truth for all patients. Based on utility theory, expected utility (EU) is a potential metric that helps demonstrate improvement in screening performance due to a rule-out device using large retrospective datasets.
    \item We applied EU to a recent study that used a large retrospective mammography screening dataset from the U.S. That study reported an improvement in specificity and decrease in sensitivity when using their AI as a rule-out device retrospectively. In terms of EU, we cannot conclude a significant improvement when the AI is used as a rule-out device.
    \item We applied the method to a European study which only reported recall rates and cancer detection rates. Since there is no established EU baseline value in European mammography screening workflow, we estimated the EU baseline using data from previous literature. We cannot conclude a significant improvement when the AI is used as a rule-out device for the European study.
    \item In this work, we investigated the use of EU to evaluate rule-out devices using large retrospective datasets. This metric, used with retrospective clinical data, could be used as supporting evidence for rule-out devices.
\end{itemize}

\newpage
\section{Introduction}
\label{introduction}
Because of its impact on breast cancer mortality and morbidity ~\cite{Karssemeijer_2009, Lehman_2017}, breast cancer screening has achieved widespread adoption despite the cost, exposure to ionizing radiation, and rates of error in the screening process. The result is approximately 40 million mammograms generated annually in the United States alone, representing a substantial workload for radiologists to interpret. In the U.S., recent recommendations for earlier screening~\cite{USPSTF-mammo-age} and reported shortages of radiologists~\cite{ACR-radiologist-shortage, RSNA-radiologist-shortage} make meeting this need even more challenging.

This has created an intense interest in using artificial intelligence (AI) algorithms to rule-out the need for radiologist interpretation of some mammograms. These algorithms (considered to be software as medical devices) are intended to autonomously remove patient images from radiologists’ reading queue, when they are deemed free of disease. In this way, they could potentially make a substantial reduction in the workload associated with screening mammography.

Ideally, a rule-out device would only exclude actually negative (i.e. signal-absent) cases from review.  This could have a beneficial impact on screening performance because of the potential elimination of cases that would otherwise lead to false-recalled outcomes.  However, if the rule-out device mistakenly removes actually positive (i.e. signal-present) cases from review, then it can potentially create false-negative outcomes that might be correctly recalled by radiologist if they were not ruled out by the device. Thus, the rule-out device has the potential for both positive and negative effects on screening performance.  This can complicate the evaluation of these devices.

Rule-out devices are often evaluated by a retrospective simulation in which the rule-out device is applied retrospectively to images that have already been read by radiologists as part of screening practice ~\cite{Rodriguez_Ruiz_2019, yala2019,Larsen2022_euro,Dembrower2020-af}. The effect of the rule-out device is simulated by combining the AI results and the reading results. These studies rely on an assumption of consistent behavior, meaning that radiologists’ screening interpretations on the cases not ruled out by the device are the same as in the clinical data. Performance (e.g. sensitivity and specificity) of the simulated combination (rule-out device and radiologist reading of the remaining cases) can be compared to the radiologist performance without the rule-out device as a reference. However, in a simulation like this, sensitivity of the combination must be the same or lower than the radiologist, representing any detected cancers that are ruled-out by the device.  And conversely, specificity must be the same or higher, reflecting any false-recalled cases ruled-out by the device. Thus, the comparison of the overall performance with and without the device is almost guaranteed to be ambiguous in terms of the sensitivity and specificity, the common measures of screening performance. The purpose of this work is to describe alternative performance measures that may resolve this performance ambiguity.

We evaluate two alternative approaches for characterizing rule-out device performance in this work. The first uses positive predictive value (PPV) and negative predictive value (NPV) as alternatives to sensitivity and specificity. We demonstrate that the PPV and NPV criteria are more inclusive and therefore more statistically powerful than sensitivity and specificity.  The second approach uses an expected utility (EU) measure~\cite{abbey2010, Abbey2012} to quantify the trade-off between sensitivity and specificity. By comparing expected utility for different screening workflows, we can determine whether the performance of combined decision making represents an improvement over the reference. 

\section{Methods}
\label{method}
\begin{figure*}[h!]
  \includegraphics[width=0.95\linewidth]{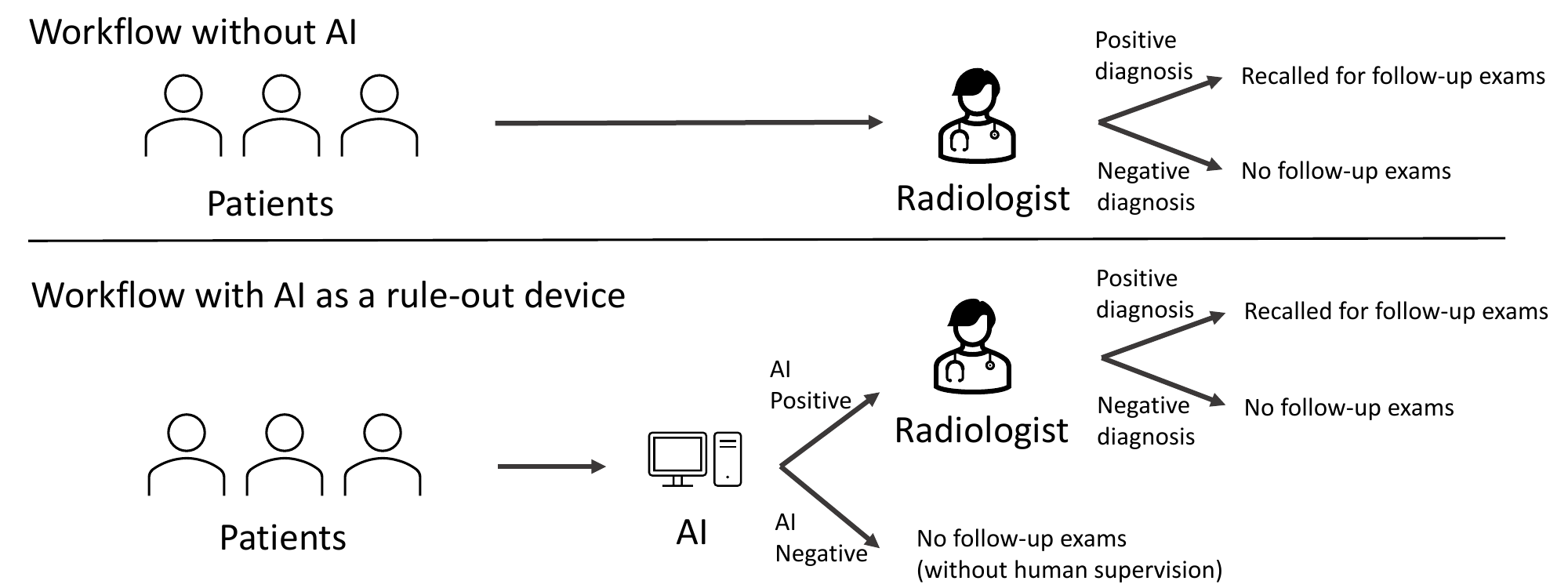}
  \centering
  \caption{Comparison between the standard-of-care workflow without an AI algorithm (top) and a workflow in which an AI is used as a rule-out device (bottom), where AI-negative patient cases are autonomously removed from the radiologist's reading queue.}
  \label{fig:workflow}
\end{figure*}

Figure \ref{fig:workflow} shows a standard reading workflow (top) and the proposed workflow including a rule-out device (bottom).  The device calls some patients negative without the need for reading by a radiologist. With retrospective clinical data, the algorithm can be applied to the images to partition the data into for-reading and rule-out subsets as a simulation of what might happen if the algorithm were applied prospectively. A key to this simulation is the assumption of consistent behavior, which allows the investigators to infer the performance impacts of the rule-out device.  This is the approach taken in the primary publications for the studies we analyze in this work ~\cite{yala2019,Larsen2022_euro}.  

In this section, we review two general evaluation frameworks (PPV/NPV and EU) which can be applied to compare the screening performance between radiologist-with-rule-out-device (which we call alternative test A) and radiologist-without-device workflows (which we call reference test B). 

\subsection{Positive and negative predictive values}
\label{predvalues}

\begin{figure}[h!]
\includegraphics[width=0.75\linewidth]{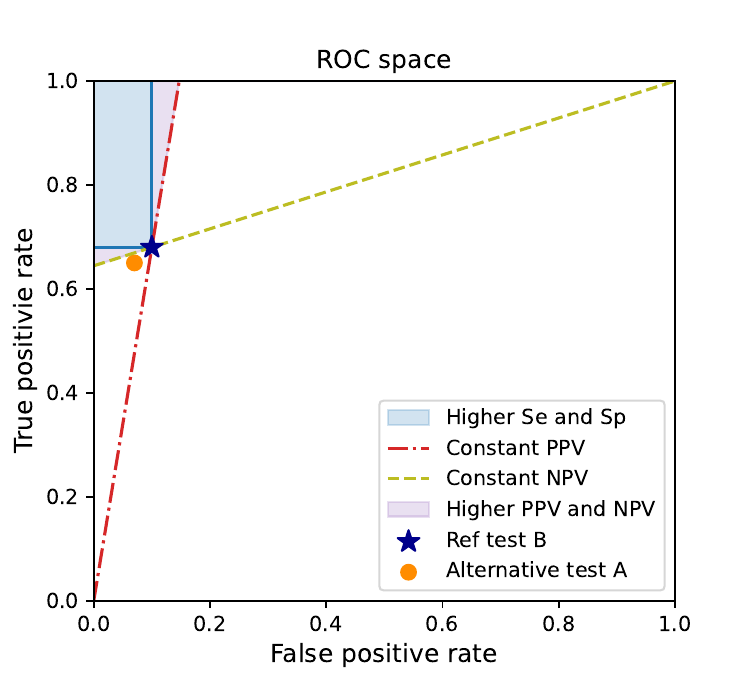}
\centering
\caption{The performance of a hypothetical alternative test A (orange dot) and that of a hypothetical reference test B (blue star) in the ROC space. If the true-positive and false-positive rates of test A fall in the blue-shaded region, test A has a better sensitivity (Se) and specificity (Sp), as well as a higher positive predictive value (PPV) and negative predictive value (NPV), than test B. If they fall in the purple-shaded region below (or on the right side of) the blue area, test A has a lower sensitivity (or specificity) but still has a higher PPV and NPV than test B. Because the purple-shaded region has a larger area than the blue region, using NPV and PPV is statistically more powerful than using sensitivity and specificity.  In this figure, test A cannot be considered superior by NPV/PPV because it is not in the blue or purple regions.}
\label{fig:roc_ppv}
\end{figure}

In general, if test A has a higher sensitivity and higher specificity than test B, then the performance of test A is superior to that of test B. Graphically in the receiver operating characteristic (ROC) space, this is equivalent to a point located within the blue upper-left region in Fig.~\ref{fig:roc_ppv} defined by the ($R_\text{FP,B}$, $R_\text{TP,B}$) point of test B (blue star in Fig.~\ref{fig:roc_ppv}). 

Biggerstaff~\cite{biggerstaff2000} demonstrated that test A may have a higher positive predictive value (PPV) and negative predictive value (NPV) than test B, even if test A has a lower sensitivity or specificity. By definition, PPV is the probability of having a signal-present case given that the test result is positive i.e. $P(D|\texttt{+})$. Via Bayes' theorem, PPV is directly related to the positive likelihood ratio $\rho_{\texttt{+}}$~\cite{biggerstaff2000}.
\begin{align}
    \label{eq:PPV}
      \text{PPV} \equiv P(D|\texttt{+}) = \frac{\rho_{\texttt{+}}}{\rho_{\texttt{+}} + Q_\pi}
\end{align}
where $Q_\pi \equiv (1-\pi)/\pi$, and $\pi$ is the prevalence of the disease of interest. Because $\rho_{\texttt{+}}$ is equal to the true-positive rate divided by the false-positive rate ($R_\text{TP}/R_\text{FP}$), $\rho_{\texttt{+}}$ represents the slope of a line connecting the ($R_\text{FP}, R_\text{TP}$) point of a diagnostic test to (0,0) in the ROC space (the red dash-dotted line in Fig.~\ref{fig:roc_ppv}). Hence, all ($R_\text{FP}, R_\text{TP}$) points along this line have a constant PPV. Since Eq.~\ref{eq:PPV} monotonically increases with $\rho_{\texttt{+}}$, all points on the left of the line have a higher PPV. Similarly, Biggerstaff~\cite{biggerstaff2000} also related NPV to the negative likelihood ratio $\rho_{\texttt{-}}$.
\begin{align}
    \label{eq:NPV}
    \text{NPV} \equiv P(ND|\texttt{-}) = \frac{Q_\pi}{\rho_{\texttt{-}} + Q_\pi}
\end{align}
where $\rho_{\texttt{-}} = (1-R_\text{TP})/(1-R_\text{FP})$ is, by definition, the slope of a line connecting the operating point of a diagnostic test to (1,1) (the olive-colored dashed line in Fig.~\ref{fig:roc_ppv}). Since Eq.~\ref{eq:NPV} monotonically increases as $\rho_{\texttt{-}}$ decreases, all points above the line have a higher NPV. If the ($R_\text{FP,A}, R_\text{TP,A}$) point of diagnostic test~A is located in the purple region of Fig.~\ref{fig:roc_ppv}, then test~A has a superior performance in terms of PPV and NPV than test~B.  In this scenario, Egan~\cite[Section~2.6.3]{Eganbook} demonstrated that test A must exist on a higher ROC curve than test B if those ROC curves are proper with a monotonically decreasing slope.

The "simulated" radiologist-with-rule-out-device workflow is equivalent to combining the radiologist-without-device test and AI test using a believe-the-negative approach. \cite{marshall} shows that the resulting negative predictive value (NPV) could either increase or decrease depending on the association between the two tests. \cite{Obuchowski} also suggests examine PPV and NPV in the rule-out scenario. When comparing the screening performance of a ``simulated'' radiologist-with-rule-out-device workflow (alternative test A) to that of a radiologist-without-device workflow (reference test B), the ``simulated'' operating point of test A is expected to have a lower false-positive rate because the device helps exclude signal-absent cases that would have been mistakenly diagnosed as positive by the radiologist. The test A operating point is guaranteed to have a greater PPV, if the rule-out device performs better than guessing, $R_\text{TP,d}-R_\text{FP,d}>0$, where the subscript $\text{d}$ represents the AI standalone performance. To determine whether test A also results in a greater NPV, it is necessary to have the true-negative rate and false-negative rate. That means the ground truth of the negative cases by both the device and radiologist is needed. This is often very difficult and costly in a screening setting since non-recalled patients usually do not receive follow-up exams. Therefore, the use of PPV/NPV may not be practical in evaluating the overall screening performance with a rule-out device for cancer screening.

\subsection{Expected utility}
\label{ERUsection}
Abbey et al.~\cite{abbey2010} showed how utility theory can be used to compare the performance of two tests, and similar utility-based approaches were developed by Baker et al.~\cite{Baker2009-zs}. The utility of a test can be expressed as
\begin{align}
\label{eq:u}
U = U_\text{TP}P(\text{TP}) + U_\text{FP}P(\text{FP}) + U_\text{TN}P(\text{TN}) + U_\text{FN}P(\text{FN})
\end{align}
where $P(\cdot)$ represents the probability of the corresponding outcome (true-positive TP, false-positive FP, true-negative TN, and false-negative FN), and $U(\cdot)$ represents the utility of each outcome. The probability of each outcome depends on the sensitivity and specificity as well as disease prevalence $\pi$. The probability of each outcome is given by $P(\text{TP}) = R_\text{TP} \pi$, $P(\text{FN}) = (1-R_\text{TP})\pi$, $P(\text{TN}) = (1-R_\text{FP})(1-\pi)$, and $P(\text{FP}) = R_\text{FP}(1-\pi)$. Rearranging the terms in Eq.~\ref{eq:u} gives the following expression for iso-utility.
\begin{align}
    \label{eq:rearrange_u}
    R_\text{TP} = \frac{(U_\text{TN}-U_\text{FP})(1-\pi)}{(U_\text{TP}-U_\text{FN})\pi}R_\text{FP} + \frac{U - U_\text{FN}\pi - U_\text{TN}(1-\pi)}{(U_\text{TP}-U_\text{FN})\pi}
\end{align}
Assuming the utility of each outcome is independent of the rate of each outcome, the iso-utility curve is a set of straight lines with a positive slope (assuming a correct decision has greater utility than an incorrect decision) in the ROC space~\cite{abbey2010}. Moreover, the expected utility $U$ is maximum when the tangent of an operating point along a ROC curve is equal to the slope of the iso-utility line.

Wagner et al.~\cite{wagner_2004} defined relative utility $U_\text{Rel}$ as follows.
\begin{align}
    \label{eq:relative_u}
    U_\text{Rel} =  \frac{(U_\text{TP}-U_\text{FN})}{(U_\text{TN}-U_\text{FP})}
\end{align}
Since the utility of a correct decision is assumed to be higher than that of an incorrect decision, both numerator and denominator are positive. Relative utility is essentially the trade-off between identifying signal-present patients and identifying signal-absent patients. Hence, for cancer screening, relative utility is much greater than 1. Based on Eq.~\ref{eq:rearrange_u}, the slope of the tangent line at the optimal point on an ROC curve is inversely proportional to the relative utility.
\begin{align}
    \label{eq:roctangent}
    \text{slope}_{\text{opt}} = \frac{Q_\pi}{U_\text{Rel}}
\end{align}
and the iso-utility equation (Eq.~\ref{eq:rearrange_u}) can be re-expressed as 
\begin{align}
    \label{eq:rearrange_u2}
    R_\text{TP} = \text{slope}_{\text{opt}} \times R_\text{FP} + c
\end{align}
where $c$ is the y-intercept and 
\begin{align}
    \label{eq:iui_eu}
    c = \frac{U - U_\text{FN}\pi - U_\text{TN}(1-\pi)}{(U_\text{TP}-U_\text{FN})\pi}
\end{align}
We can see that $c$ is linearly related to the expected utility.
We can also write the equation \ref{eq:rearrange_u} into recall rate ($R_R$) and detection rate ($R_D$) space by substituting the two equations below into equation \ref{eq:rearrange_u2}:
\begin{align}
    \label{eq:rd}
    R_\text{D} = \pi R_\text{TP} 
\end{align}
\begin{align}
    \label{eq:rr}
    R_\text{R} = \pi R_\text{TP} + (1-\pi)R_\text{FP}
\end{align}
 And rearrange the terms to obtain
 \begin{align}
    \label{eq:rearrange_u3}
    R_\text{D} = \frac{1}{1+U_\text{Rel}}R_\text{R} + c_2.
\end{align}
where $c_2$ is linearly related to the expected utility. 

\begin{figure}[h!]
  \includegraphics[width=0.75\linewidth]{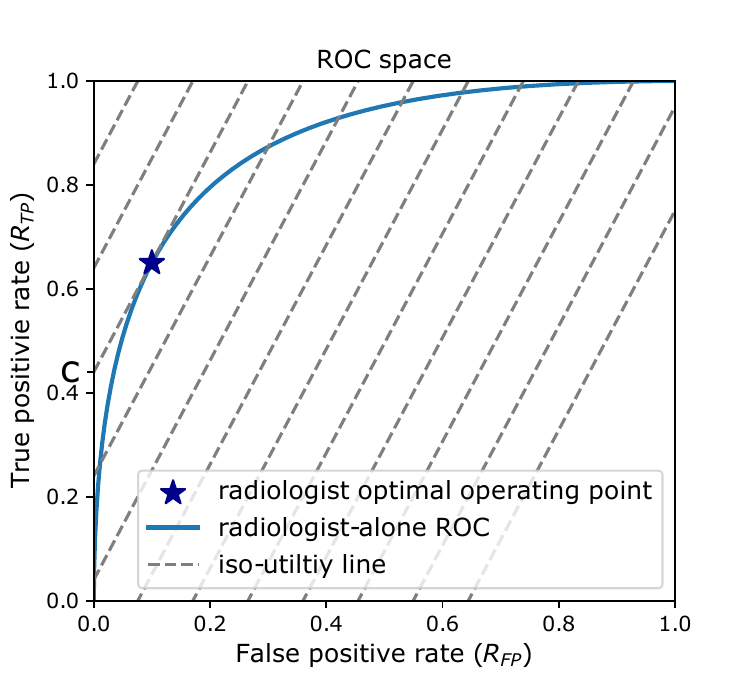}
  \centering
  \caption{The blue line represents a hypothetical ROC curve in a radiologist-without-device workflow. The gray dashed lines represent a set of iso-utility lines. The slope of the iso-utility lines is $Q_\pi/U_\text{Rel}$. Iso-utility lines with a higher y-intercept (IUI) represent a higher expected utility. The blue star shows an example optimal point along the ROC curve, and the tangent at the optimal point (maximum utility) is equal to the slope of the iso-utility line. The intercept $c$ is given 
  by Equation~\ref{eq:y-in_ROC}.}
  \label{fig:tangent}
\end{figure}

The value of relative utility is often unknown. However, as Abbey et al.~\cite{Abbey2012} noted, assuming that an experienced radiologist operates at the optimal point, one could estimate the relative utility from the tangent line at the radiologist's operating point along the ROC curve (blue star in Fig.~\ref{fig:tangent}). Using a large dataset from the DMIST study in the U.S.~\cite{Pisano_2005}, Abbey et al. estimated a relative utility of $162\pm14\%$ for a standard-of-care screening mammography workflow. That is, the utility of identifying a signal-present patient is 162 times more important than that of identifying a signal-absent patient. In this work, 162 is used as the baseline relative utility value for screening mammography in the radiologist-without-device workflow in the U.S.

\begin{figure}[h!]
  \includegraphics[width=0.8\linewidth]{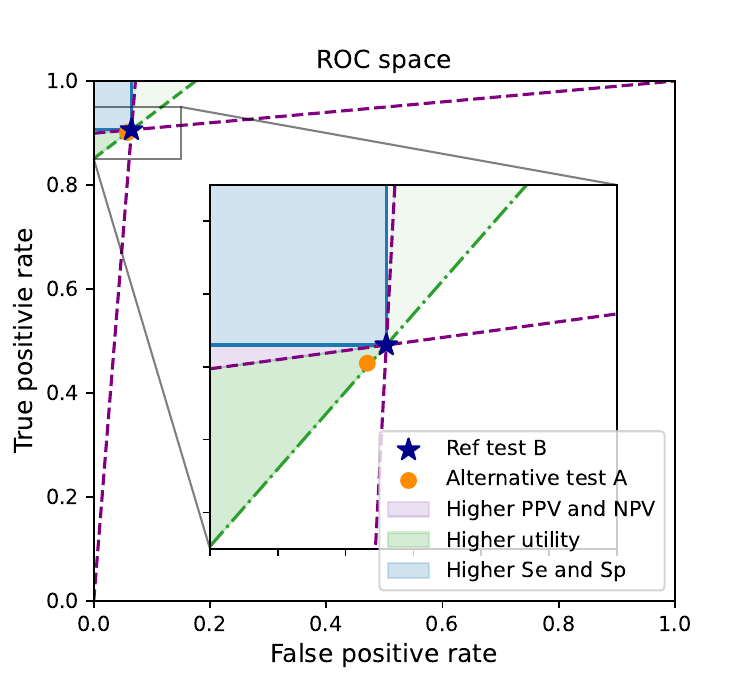}
  \centering
  \caption{Regions where a ``simulated'' radiologist with a rule-out device (alternative test A) has a superior screening performance compared to that of a radiologist-without-device workflow (reference test B). The blue-shaded region indicates the alternative test A has a better sensitivity (Se) and specificity (Sp) than the reference test B. The purple-shaded region indicates that test A has either a lower sensitivity or specificity but a higher positive predictive value and negative predictive value. The green-shaded region indicates test A has a higher utility than test B. The relative utility along the green dashed line was estimated to be 162 by Abbey et al.~\cite{Abbey2012} for U.S. screening mammography. The operating points of test A (orange dot) and test B (blue star) as well as the disease prevalence are based on the paper by Yala et al.~\cite{yala2019}.}
  \label{fig:roc}
\end{figure}

Using the baseline relative utility, one can determine whether an alternative test A has a better performance than a reference test B with an iso-utility line. If test A has an operating point (orange dot in Fig.~\ref{fig:roc}) that is above the iso-utility line of test B (green dashed line in Fig.~\ref{fig:roc}), then the expected utility of test A is higher. In other words, test A has a higher expected utility if the y-intercept of test A's iso-utility line is higher than the y-intercept of test B's iso-utility line. The green-shaded region in Fig.~\ref{fig:roc} shows the region where an alternative test A is superior to the reference test B in terms of expected utility.

Given a $(R_{TP}, R_{FP})$ pair, the y-intercept of its iso-utility line in the ROC space can be estimated by 
\begin{align}
    \label{eq:y-in_ROC}
    c = R_{TP} - \frac{Q_\pi}{U_{Rel}}R_{FP}
\end{align}
The quantity $c$ is unitless and linearly related to the expected utility as shown in equation \ref{eq:iui_eu}, so we can use it as a proxy for the expected utility in equation \ref{eq:u} and as a metric for comparing the performance of tests given the true relative utility.  For convenience, we name $c$ the Iso-Utility Intercept or IUI. 95\% confidence intervals (CI) of the IUI are calculated via bootstrapping from 5000 samples.

Alternatively, given a $(R_D,R_R)$ pair, the y-intercept of its iso-utility line in the detection/recall space can be estimated by
\begin{align}
    \label{eq:y-in_D_R}
    c_2 = R_{D} - \frac{1}{1+U_{Rel}}R_{R}
\end{align}
Because $c_2$ is also linearly related to EU, we can use it as a proxy to EU. Therefore, $c_2$ could be used as a metric for the performance of the two tests given the true relative utility. Measured detection and recall rates are common in real-world studies which makes this method advantageous. For convenience, we name $c_2$ the Detection Iso-Utility Intercept or DIUI.

For both PPV/NPV and EU methods, we calculated the probability that the metric from a radiologist-with-device workflow is greater than that from a without-device workflow, as well as the CI on the probability, via bootstrapping with 5000 samples.

\subsection{Two recent publications}
\label{datasets}
In this work, we apply both PPV/NPV and expected utility methods on two retrospective ``simulated'' studies. First is a U.S. study by Yala et al.~\cite{yala2019}, and second is a European study by Larsen et al.~\cite{Larsen2022_euro}.  This section summarizes the two publications.

Yala et al.~\cite{yala2019} retrospectively applied a deep-learning AI algorithm as a potential rule-out device on a large U.S. mammography dataset. The AI was trained using 212,272 mammograms from 56,831 patients and was evaluated using 26,540 mammograms from 7,176 patients with a prevalence of 0.7\%. The AI-alone ROC curve has an area under the ROC curve (AUROC) of 0.85 [95\% CI: 0.80, 0.85]. All data was retrospectively collected at a large medical center in the U.S. in which one radiologist read (without any AI-enabled aiding or rule-out devices) and determined whether a patient should be followed up. The radiologist-without-device workflow has a sensitivity of 90.6\% [95\% CI: 86.6\%, 94.7\%] and a specificity of 93.5\% [95\% CI: 86.0\%, 93.9\%]. The authors then retrospectively ``simulated'' a radiologist-with-rule-out-device workflow in which a patient would only get a follow-up exam when the patient was diagnosed as positive by the radiologist and also labeled as positive by the AI. The radiologist-with-rule-out-device workflow has a sensitivity of 90.1\% [95\% CI: 86.0\%, 94.3\%] and a specificity of 94.2\% [95\% CI: 94.0\%, 94.6\%], and the AI would have excluded 19.3\% of cases. The authors also provided the sensitivity and specificity values of the radiologist-with-rule-out-device workflow as a function of the fraction of excluded mammograms in Table E2 of their appendix. When applying the expected utility method to this study, we use the relative utility of 162 as our baseline value~\cite{Abbey2012}. 

Larsen et al.~\cite{Larsen2022_euro} investigated the potential clinical impacts that a rule-out device could bring to the standard-of-care, double-reading mammography screening. An AI-enabled lesion detection system developed for mammography screening (with an AUROC of 0.84 [95\% CI: 0.82, 0.86]~\cite{Rodriguez2019}) was used to retrospectively analyze 122,969 mammograms from 47,877 patients in Norway. From the standard-of-care workflow, the recall rate and cancer detection rates are 3.2\% and 0.61\% respectively. Using the AI score and ratings from two radiologists, the authors defined 11 hypothetical workflows and calculated their recall rates, cancer detection rates, and sensitivities. No specificity values were reported. Due to the difference in single- versus double-reading between the U.S. and Europe, we focus on Scenarios 3-5 (defined in Table 1 in~\cite{Larsen2022_euro}) in which the AI is hypothetically used to pre-screen cases before they enter radiologists' worklist. These three workflows correspond to the AI ruling out 30\%, 50\%, and 70\% of the total patients, and their estimated recall rates (and cancer detection rates) are 2.6\% (0.6\%), 2.1\% (0.58\%), and 1.2\% (0.53\%) respectively (Table 2 in~\cite{Larsen2022_euro}).

When applying the expected utility method to European studies, we cannot use the same relative utility baseline value (162) because of the difference in radiologist training and clinical practice. Several studies found that European mammography radiologists have a lower recall rate compared to the U.S. radiologists~\cite{Smith_Bindman_2003, Fletcher_2005, Domingo_2015}. Furthermore, double-reading for mammography screening is very common in Europe~\cite{Taylor_Phillips_2018} compared to single reading in the U.S.; mammograms in Europe are reviewed by two radiologists who must reach a consensus if their initial decisions disagree. Therefore, the ROC curves and reader thresholds are expected to differ significantly between radiologists in Europe and those in the U.S., and the trade-off between signal-present and signal-absent patients is expected to be different. Hence, the baseline relative utility of 162 is not applicable to European studies, and we will use previously published result~\cite{otten_elat} to estimate the relative utility of European double-reading.

\section{Results}
\label{analysis}
We present the results of applying the PPV/NPV and expected utility methods in Section~\ref{method} to two studies: a U.S. study by Yala et al.~\cite{yala2019} and a European study by Larsen et al.~\cite{Larsen2022_euro}. 

\subsection{U.S. mammography screening}
Using the reported values of prevalence, sensitivity, and specificity for the radiologist-without-device workflow in  Yala et al.~\cite{yala2019}, we calculate PPV (Eq.~\ref{eq:PPV}) and NPV (Eq.~\ref{eq:NPV}) to be 9.18\% and 99.9\% respectively. Figure~\ref{fig:roc} shows the lines of constant PPV and NPV (dashed purple lines) at the sensitivity and specificity of the radiologist-without-device workflow. Since the reported sensitivity and specificity of the radiologist-with-rule-out-device workflow is located outside the purple region, the screening performance of the radiologist-with-rule-out-device workflow is not superior to that of the radiologist-without-device workflow in terms of PPV/NPV. The probability of the radiologist-without-rule-out-device workflow having a greater PPV/NPV value is 36\%. 

With the same set of prevalence, sensitivity, and specificity, the radiologist-without-device workflow has an IUI of 0.85 [95\% CI: 0.806, 0.89] in the ROC space, and the radiologist-with-device workflow has an IUI of 0.851 [95\% CI: 0.806, 0.895]. Since the radiologist-with-device IUI estimate is above the one without the device, the radiologist-with-device operating point is within the region of higher expected utility (green area in Fig.~\ref{fig:roc}). However, because its lower bound of the 95\% CI is below the IUI of radiologist-without-device workflow, we cannot conclude with statistical certainty that their AI provides a significantly better utility if used as a rule-out device at the operating threshold. The probability of the radiologist-with-device workflow having a greater utility is 72.1\%. On the other hand, the change in the expected utility is near negligible, implying the radiologist-with-device workflow could maintain a similar expected utility while reducing workload.

\begin{figure}[h!]
  \includegraphics[width=0.75\linewidth]{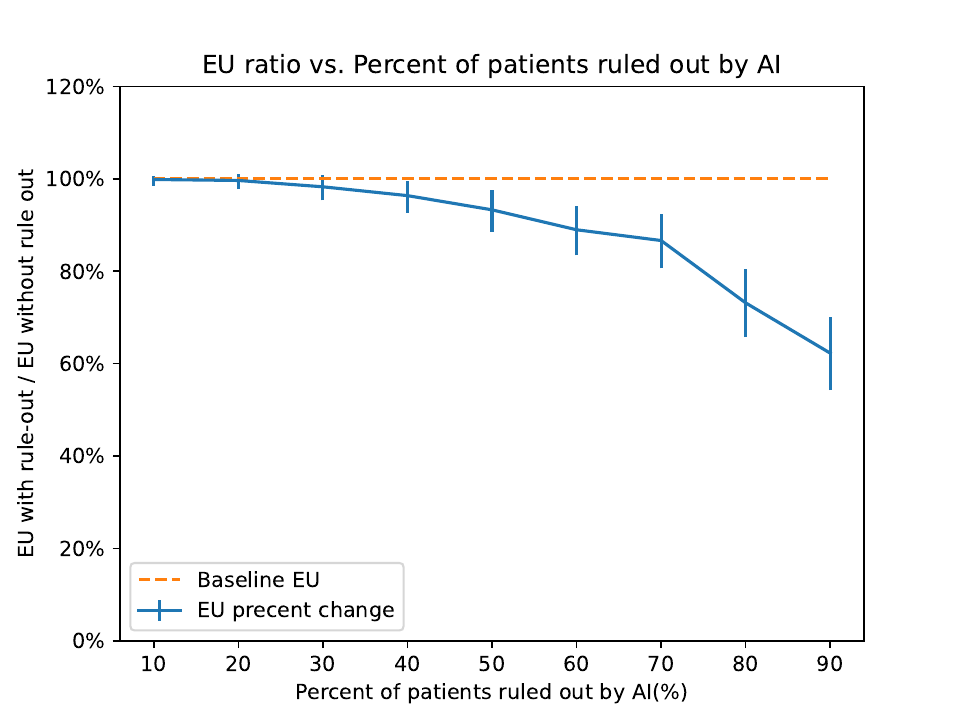}
  \centering
  \caption{Ratio of EU from a radiologist-with-device workflow to that from a without-device workflow as a function of the percentage of patients excluded by the AI algorithm, using the AI thresholds in Table E2 of the Appendix in Yala et al.~\cite{yala2019}. The blue line shows the percentage change of EU for different AI thresholds, and the vertical error bars represent the 95\% CIs. The horizontal orange dashed line represents the EU ratio of 1 assuming a baseline relative utility of 162 from Abbey et al.~\cite{Abbey2012}.}
  \label{fig:RU_US}
\end{figure}

Using other AI thresholds reported in Table E2 (Appendix of Yala et al.~\cite{yala2019}), the IUI values, their 95\% CIs, and the probabilities that the radiologist-with-device workflow has a greater utility and PPV/NPV are summarized in Table~\ref{tab:yalaresult}. The IUI values and their 95\% CIs are plotted as a function of the percentage of cases excluded by the potential rule-out device in Fig.~\ref{fig:RU_US}. Based on the available operating thresholds, none of them achieve an IUI point estimate higher than the radiologist-without-device workflow. However, the drop in expected utility is only 5\% when ruling out 30\% of patients, indicating that the drop in performance could be negligible at low rule-out fraction.

\begin{table}[ht!]
\begin{center}       
\begin{tabular}{|r|c|c|r|r|c|c|}
\hline
\rule[-1ex]{0pt}{3.5ex} {\shortstack{Rule out \\ patient(\%)}} & IUI   & \rule[-1ex]{0pt}{3.5ex} \shortstack{95\% CI\\ on IUI}   &\multicolumn{1}{c|}{  \shortstack{ P(IUI\textgreater{}\\baseline)}} & \multicolumn{1}{c|}{\shortstack{P(PPV,NPV$>$ \\ baseline)}} & Se & Sp \\ \hline
baseline, 0\%    & 0.85      & (0.806,0.890)    & -        & -        & 90.6\%    & 93.5\%    \\ \hline
10\%             & 0.849     & (0.803,0.890)    & 36.5\%   &36.4\%    & 90.1\%    & 93.9\%    \\ \hline
20\%             & 0.847     & (0.801,0.888)    & 40.9\%   &13.2\%    & 89.5\%    & 94.3\%    \\ \hline
30\%             & 0.835     & (0.787,0.882)    & 12.5\%   &0.6\%     & 88.0\%    & 94.8\%    \\ \hline
40\%             & 0.819     & (0.768,0.865)    & 1.6\%    &0\%       & 85.9\%    & 95.4\%    \\ \hline
50\%             & 0.793     & (0.737,0.845)    & 0.0\%    &0\%       & 82.7\%    & 96.0\%    \\ \hline
60\%             & 0.756     & (0.696,0.813)    & 0.0\%    &0\%       & 78.5\%    & 96.6\%    \\ \hline
70\%             & 0.736     & (0.676,0.796)    & 0.0\%    &0\%       & 75.9\%    & 97.3\%    \\ \hline
80\%             & 0.622     & (0.553,0.690)    & 0.0\%    &0\%       & 63.9\%    & 98.0\%    \\ \hline
90\%             & 0.529     & (0.460,0.599)    & 0.0\%    &0\%       & 53.9\%    & 98.8\%    \\ \hline
\end{tabular}
\end{center}
\caption{Table summarizing iso-utility intercept as a function of patient rule-out percentage, sensitivity, and specificity reported in Table E2 of the Appendix of Yala et al.~\cite{yala2019}. From left to right, the columns are the percentage of total patients ruled out by the AI algorithm~\cite{yala2019}, the iso-utility intercept (IUI), its 95\% CI, the probability (P) that the IUI of the with-device workflow is greater than without-device baseline (assuming a relative utility of 162), the probability of PPV and NPV being greater than baseline, and the sensitivity and specificity reported in ~\cite{yala2019}.}
\label{tab:yalaresult}
\end{table} 

\subsection{European mammography screening}\label{analysis-euro}
Although sensitivities for all scenarios were reported (Fig. 2 in Larsen et al.~\cite{Larsen2022_euro}), the authors did not provide the estimated specificities.  With only sensitivity, we cannot calculate NPVs, and the PPV/NPV method cannot be applied.

With the reported recall rates and cancer detection rates, we can apply the EU method (Eq.~\ref{eq:y-in_D_R}) if we know the baseline relative utility. However, no previous study on European double-reading mammography screening established a baseline relative utility. Therefore, we first estimated the baseline utility for double-reading European mammography screening. We took the recall and detection rate data from Otten et al.~\cite{otten_elat} and interpolated the data point using cubic spline. We then calculated the derivative of the interpolated curve at a recall rate of 3.2\%, the same as the Larsen et al. reported recall rate. From the slope of the curve at a recall rate of 3.2\%, we obtained a relative utility of 111. This number is significantly lower than the US baseline, but it aligns with published literature that European double-reading screening operates at lower recall rates~\cite{Smith_Bindman_2003, Fletcher_2005, Domingo_2015}. We use the 111 number as the baseline utility for the European study.

\begin{figure}[h!]
  \includegraphics[width=0.75\linewidth]{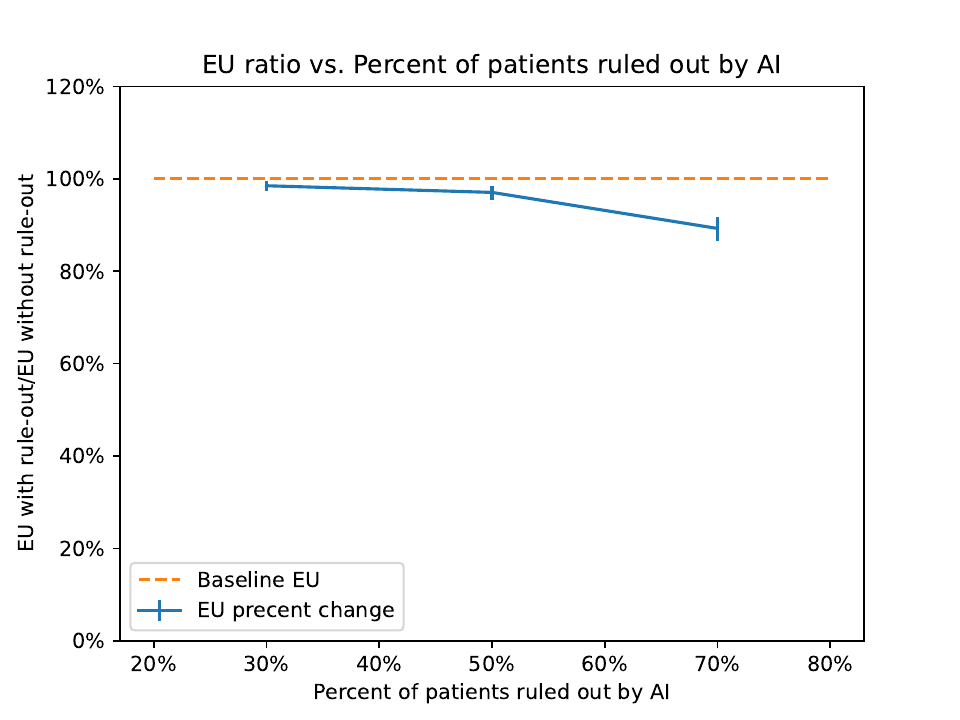}
  \centering
  \caption{Ratio of EU from a radiologist-with-device workflow to that from a without-device workflow
as a function of the percentage of patients excluded by the AI algorithm, using data from Larsen et al.~\cite{Larsen2022_euro}. The blue line shows the percentage change of EU for different percentages of patients ruled out (AI decision threshold) when using a baseline relative utility of 111. The vertical error bar represents the 95\% CI.}
  \label{fig:eru_europe}
\end{figure}

A summary of the DIUI values in detection/recall space and their 95\% CIs of different scenarios are shown in Table~\ref{tab:euro_larsen} and Fig.~\ref{fig:eru_europe}. All rule-out fractions result in a drop in the DIUI (and hence expected utility), but the drop in DIUI is still only around 10\% even at 70\% rule-out fraction. This suggests a relatively small cost in terms of expected utility could result in a huge reduction in reading volume.

\begin{table}[ht]
\begin{center}       
\begin{tabular}{|r|c|c|c|c|c|}
\hline
\rule[-1ex]{0pt}{3.5ex}{\shortstack{Rule-out \\patient (\%)}} & 
 \shortstack{DIUI\\($10^{-3})$} & 
 \shortstack{95\% CI of DIUI\\($10^{-3}$)} & 
 \shortstack{p(EU \textgreater{}\\baseline) (\%)} & \shortstack{Recall\\rate (\%)} &\shortstack{ Detection\\rate (\%)} \\ \hline
baseline, 0 & 5.83 & (5.40,6.27)  & -      & 3.2   & 0.61   \\ \hline
30          & 5.74 & (5.31,6.17)  & 0.18   & 2.6   & 0.60   \\ \hline
50          & 5.66 & (5.24,6.08)  & 0      & 2.1   & 0.58   \\ \hline
70          & 5.21 & (4.81,5.60)  & 0      & 1.2   & 0.53   \\ \hline
\end{tabular}
\end{center}
\caption{Table summarizing the result of applying EU methods to Scenarios 3-5 defined in Larsen et al.~\cite{Larsen2022_euro}. From left to right, the columns are the percentage of total patients ruled out by the AI if used as a rule-out device, the detection iso-utility intercept (DIUI) in detection/recall space, the corresponding 95\% CI, the probability that the expected utility of the rule-out scenario is higher than that of the without-device baseline, and the recall and detection rates reported in Larsen et al.~\cite{Larsen2022_euro}.}
\label{tab:euro_larsen}
\end{table}

\section{Discussion}
\label{discussion}


Using two recently published retrospective ``simulated'' analyses (one in the U.S. and one in Europe), we demonstrated how PPV/NPV and EU can be applied to determine whether the ``simulated'' radiologist-with-rule-out-device workflow has a superior screening performance than the standard-of-care, radiologist-without-device workflow. For the U.S. study, despite the reported improvement in sensitivity and specificity, neither PPV/NPV nor EU supports the claim of a significant improvement for any of the algorithm thresholds reported. However, we found that the drop in expected utility is not significant until ruling out more than 30\% patients. 

The fact that neither PPV/NPV nor EU concludes a significant improvement in screening performance compared to standard-of-care workflow for both studies reflects the challenges in demonstrating device effectiveness when an AI is used as a rule-out device. Because there is always a trade-off between sensitivity and specificity, they are ambiguous metrics when, by nature of the ``simulated'' retrospective study design, sensitivity can only go down and specificity can only go up. Metrics such as PPV/NPV and EU is more appropriate than sensitivity and specificity for this type of ``simulated'' retrospective study design to demonstrate an unambiguous improvement in screening performance when such an improvement exists. 

One main advantage of both PPV/NPV and EU is that they are readily applied to any large observational clinical study. These observational studies may collect only radiologists' binary decisions and lack finer radiologist ratings that we may measure in a controlled, prevalence-enriched reader study~\cite{FDA-clinical-assessment}. However, these observational studies are collected in real clinical environments and have large numbers of patients and image readings. The use of PPV/NPV and EU can reduce ambiguity in assessing rule-out device performance from large observational clinical studies.

As demonstrated in the European example in Sec.~\ref{analysis-euro}, another benefit of using EU instead of PPV/NPV is that EU can be calculated using only recall rate and cancer-detection rate. Given the low prevalence in a screening setting, a majority of patient cases in large clinical studies are not recalled by radiologists, and their ground truth cannot be determined. Hence, it is clinically impractical to estimate the specificity of these large studies. Using the recall rate and cancer detection rate in the EU method eliminates the needs of knowing the false-positive rates, which is less burdensome compared to using PPV/NPV.

In addition, with the EU method, we showed how the expected utility of a radiologist-with-rule-out-device workflow will change as a function of patient ruled-out fraction and directly compared it to the EU of a radiologist-without-rule-out-device workflow. We could see at what fraction of patients ruled out, that the drop in expected utility is significant enough and outweighs the benefit of the reduced reading volume.

As for their limitations, neither of the two methods provide insights on the full radiologist-with-rule-out-device ROC curve or AUROC. Both PPV/NPV and EU methods compare the operating points before and after the device is hypothetically included in the workflow and assume that the radiologists do not adjust their decision-making threshold when the AI is used as a rule-out device. In reality, due to the increase in prevalence, radiologists are expected to adjust their decision thresholds to obtain a higher expected utility. We expect this adjustment in radiologist threshold and utility to increase as the rule-out fraction increases. A reader study that elicits reader confidence ratings would yield a full ROC curve which would map the reader performance at all decision thresholds from which a hypothesis test using AUROC or partial AUC can be performed. However, because the retrospectively ``simulated" operating point should lie on the full radiologist-with-device ROC curve, this simulated operating point is still valuable in providing a lower bound on the expected utility even when the radiologists adjust their operating points. In other words, the $EU_{\text{rad+AI}}>EU_{\text{rad}}$ based on a retrospective, simulated analysis is a sufficient condition but not a necessary condition for the radiologist-with-rule-out-device workflow to have a higher expected utility than the radiologist-without-device scenario. 

One disadvantage of the EU method is the need of a baseline relative utility value. This work relies on the value of 162 estimated by Abbey et al.~\cite{Abbey2012} and 111 estimated from Otten et al.~\cite{otten_elat}. Abbey et al. found that different ROC models and truthing criteria resulted in different relative utility estimates, creating a larger systematic uncertainty in addition to the statistical uncertainty. Furthermore, the estimated value did not take into account patient subgroups, including breast density, type of lesions, patients' demography information, and/or their risk levels. In reality, radiologists might choose to recall more patients with a higher risk of cancer, resulting in a higher relative utility. This problem may be overcome by estimating the relative utility baseline values for each subgroup. 

It is important to note that both PPV/NPV and EU methods can be applied, not only to mammography screening, but also to other types of screening (e.g. ultrasound breast cancer screening, lung cancer screening, etc.). In terms of utility, if the radiologist-with-rule-out-device workflow achieves an EU significantly greater than the baseline expected utility, it is sufficient to conclude that the radiologist-with-rule-out-device workflow is better compared to the radiologist-without-device scenario in terms of utility. However,  achieving a higher EU may not be sufficient to demonstrate safety and effectiveness of the rule-out device. Other factors, such as shifts in radiologists' decision thresholds and correlations between radiologists' decisions and AI outputs, should also be considered.

\section{Conclusion}
\label{Conclusion}

In principle, rule-out devices may be able to eliminate a substantial number of true-negative cases, leading to a substantial reduction in workload with no impact on screening performance.  However, ``simulated'' analysis of retrospective data suggests that this is not the case.  In published U.S. and European screening data, we see that rule-out devices do not exclusively rule out cases that would otherwise be true positives.  As a result, sensitivity may decrease (by ruling out true-positive cases) while specificity may increase (by ruling out false-positive cases). As a result, performance evaluation based on sensitivity and specificity will be ambiguous. We have evaluated PPV/NPV and Expected Utility (EU) as alternative measures of screening performance that can provide a more definitive result. 

While using PPV/NPV is more statistically powerful than using sensitivity and specificity,   the hypothesized performance with the rule-out device in the U.S.-based study leads to an increase in PPV with a concomitant decrease in NPV, and is therefore also ambiguous. For the European study, NPV cannot be evaluated because false-negative cases are not identified.

We find EU to be useful in this scenario. It can be computed from sensitivity and specificity, or from recall and detection rates, it is more powerful than using PPV/NPV, and it provides a single measure of performance for comparing the impact of a rule-out device to performance without the rule-out device. Computing the expected utility requires some sense of the relative utility for the screening mammography program, which may be different in different programs, but it resolves the ambiguity in performance measures.

As a secondary finding of this work, we would note that the observed drop in EU is relatively small and not statistically significant for small rule-out fractions, suggesting that the drop in performance utility might be compensated by the benefit of reduced reading volume. 

\newpage
\listoffigures
\newpage
\listoftables

\newpage
\section*{Acknowledgement}
The authors acknowledge the Research Participation Program at the Center for Devices and Radiological Health administered by the Oak Ridge Institute for Science and Education through an interagency agreement between the U.S. Department of Energy and the U.S. Food and Drug Administration (FDA). 

The mention of commercial products, their sources, or their use in connection with material reported herein is not to be construed as either an actual or implied endorsement of such products by the Department of Health and Human Services. This is a contribution of the U.S. Food and Drug Administration and is not subject to copyright.

\newpage
\section*{Statements and Declarations}
\subsection*{Ethical considerations}
This study did not involve human or animal subjects.
\subsection*{Consent to participate}
Not applicable
\subsection*{Consent for publication}
Not applicable
\subsection*{Declaration of conflicting interests}
The authors declare that there is no conflict of interest.
\subsection*{Funding statement}
Financial support for this study was provided fully by the Center for Devices and Radiological Health (CDRH) at the U.S. Food and Drug Adminstration (FDA). 
\subsection*{Data availability}
Not applicable
\newpage
\bibliography{references}
\bibliographystyle{ref} 

\end{document}